\documentclass[pra,twocolumn,amsmath,amssymb,superscriptaddress,aps,10pt]{revtex4-1} %

\usepackage{graphicx}	
\usepackage{dcolumn}	
\usepackage{braket}	
\usepackage{comment}
\usepackage{hyperref}
\hypersetup{
    colorlinks=true,        
    linkcolor=blue,          
    citecolor=blue,        
    filecolor=magenta,      
    urlcolor=blue           
}

\def\laco{LaCoO$_3$}
\def\lsco{LaSrCoO$_4$}
\def\scof{Sr$_2$CoO$_3$F}

\usepackage{color} 				
\usepackage[normalem]{ulem} 	

\begin{document}
\title{Pressure-induced spin-state ordering in \scof}

\author{Juan Fern\'{a}ndez Afonso}
\affiliation{Institute of Solid State Physics, TU Wien, Wiedner Hauptstra{\ss}e 8, 1020 Vienna, Austria}
\author{Andrii Sotnikov}
\affiliation{Institute of Solid State Physics, TU Wien, Wiedner Hauptstra{\ss}e 8, 1020 Vienna, Austria}
\affiliation{Akhiezer Institute for Theoretical Physics, NSC KIPT, Akademichna 1, 61108 Kharkiv, Ukraine}
\author{Atsushi Hariki}
\affiliation{Institute of Solid State Physics, TU Wien, Wiedner Hauptstra{\ss}e 8, 1020 Vienna, Austria}
\author{Jan Kune\v{s}}
\affiliation{Institute of Solid State Physics, TU Wien, Wiedner Hauptstra{\ss}e 8, 1020 Vienna, Austria}
\affiliation{Institute of Physics, Czech Academy of Sciences, Na Slovance 2, 182 21 Praha 8, Czechia}

\date{\today}

\begin{abstract}
We study theoretically low-temperature phases of a recently synthesized compound \scof{} under pressure.
The analysis combining LDA+DMFT and a strong-coupling effective model points to the existence of not only normal paramagnetic and antiferromagnetic regimes, but also a spin-state ordered phase in a certain range of applied pressure and low temperature. 
This order is characterized by a checkerboard arrangement of different spin states of cobalt atoms in the lattice.
\end{abstract}

\maketitle

\section{Introduction}
Trivalent cobalt oxides with perovskite structure host a number of unusual physical phenomena resulting from the presence of several low-energy multiplets in the atomic spectrum of octahedrally coordinated Co$^{3+}$~\cite{Tanabe1954}. The parent compound LaCoO$_3$ has been attracting attention since the 1950s due to its thermally driven spin-state crossover and insulator-metal transition~\cite{Goodenough71}. Hole doping as in La$_{1-x}$Sr$_x$CoO$_3$ leads to formation of clusters with large magnetic moments and eventually to the appearance of a ferromagnetic metal with increasing hole concentration \cite{wu03,Phelan2006}. 
Ferromagnetism is observed also in strained films of LaCoO$_3$~\cite{fuchs07,fujioka13prl}, although its insulating character suggests that a different mechanism is at play. In the layered perovskite LaSrCoO$_4$ low-temperature spin glass behavior was reported~\cite{guo16}.
Compounds from the (Pr$_{1-y}$Y$_{y}$)$_{x}$Ca$_{1-x}$CoO$_3$ family exhibit a 'hidden' order at temperatures as high as 130~K, which breaks time-reversal symmetry, but does not exhibit ordered moments~\cite{tsubouchi04,hejtmanek10}. 
Understanding coupling between Co ions is essential to capture these diverse phenomena. Recent resonant inelastic x-ray scattering (RIXS) experiments on LaCoO$_3$~\cite{wang18prb} showed that not only spin exchange, but also mobile spinful excitons are important for the low-energy physics of cobaltites. 
Studies on simplified models~\cite{kunes14a,hoshino16,nasu16} as well as material specific mean-field calculations on cobaltites~\cite{afonso17,Hsu2012,yamaguchi17} uncovered a number of possible ordered phases in the vicinity of spin-state crossover including antiferromagnet (AFM), spin-state order (SSO) characterized by static checkerboard arrangement of atoms in distinct spin states, or an exciton condensate. In particular, the last one giving rise to a peculiar magnetism has been the subject of intense research recently~\cite{Khaliullin2013,Cao2014,Jain2017}.

External pressure allows experimental control of the crystal-field splitting and relative energies of multiplets on the Co site. It eventually converts a high-spin (HS) to a low-spin (LS) atomic ground state. Trivalent RCoO$_3$ (R$=$La,\,Pr,\,Y) perovskites host the LS Co$^{3+}$ already at ambient pressure and thus the spin-state crossover cannot be induced by its increase. 
In contrast, \scof{}, a recently synthesized compound~\cite{Tsujimoto11}, contains HS Co$^{3+}$ ions that order antiferromagnetically below $T_{\rm N}=323$~K
at ambient pressure~\cite{Tsujimoto12}. Unlike its layered analog LaSrCoO$_4$~\cite{guo16}, \scof{} is free from structural disorder. 
Under pressure, the ground-state configuration changes from HS to LS. At room temperature, the full conversion to LS is concluded at around 12~GPa~\cite{Tsujimoto16}. 

In this paper, we study the pressure-driven spin-state crossover in \scof{} and explore possible ordered phases by combining {\it ab initio} and mean-field numerical approaches. Our results reproduce the experimental observations of HS AFM at low pressure and LS state at high pressure. In addition, at intermediate pressure we predict the appearance of the SSO phase.
We compare the electronic structure of \scof{} with isoelectronic LaCoO$_3$ and its layered analog LaSrCoO$_4$ \cite{guo16,afonso18}.

\section{Computational method}\label{comp}
We use the multiband Hubbard model to capture the electronic correlation in the studied material. The calculations are carried out in several steps. 
Starting from density functional theory (DFT) calculations, 
the bilinear part of the Hubbard Hamiltonian is obtained by projecting the DFT Hamiltonian onto the Co-$3d$ Wannier orbitals. To incorporate the electronic correlation effects we add the local Coulomb interaction parametrized by the Slater integrals $F_0$, $F_2$, and $F_4$. Two theoretical approaches are applied for the same set of input parameters: strong coupling expansion followed by the mean-field treatment of the effective model (SC-MF) and dynamical mean-field theory (DMFT).

\subsection{LDA and tight-binding parameters}

\begin{figure}
\includegraphics[width=\columnwidth]{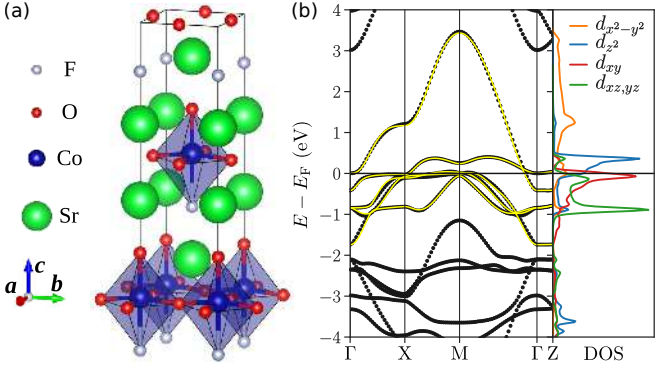}
\caption{(a) Crystal structure of \scof{}. (b) LDA band structure
plotted together with the Co-$3d$ shell 
Wannier projections (yellow)
and partial densities of states (DOS) at $P=6.4$~GPa.}
\label{structure_lda_dos}
\end{figure}

The electronic structure calculations are performed in the framework of the DFT with local density approximation (LDA)~\cite{ksdft,hklda} to the exchange correlation potential.
\scof{} consists of layers of distorted CoO$_3$F corner-sharing octahedra separated by Sr atoms; see Fig.~\ref{structure_lda_dos}(a).
%
%
The crystal parameters are taken from Ref.~\cite{Tsujimoto11}, amounting to  
$a = 3.83145$~{\AA}  and $c = 13.3201$~{\AA} at ambient pressure.
Following Ref.~\cite{Tsujimoto16}, the space group $I4/mmm$ is kept unchanged throughout the studied pressure range.
The calculations are carried out with the \textsc{wien2k} package~\cite{wien2k}. 
The muffin-tin radii are set 2.22 for Sr, 1.90 for Co, 1.63 for O, and 2.18 for F in atomic units.
The Brillouin zone was sampled with a $k$-mesh grid of the size $10 \times 10 \times 10$. 

\begin{figure}
\includegraphics[width=\columnwidth]{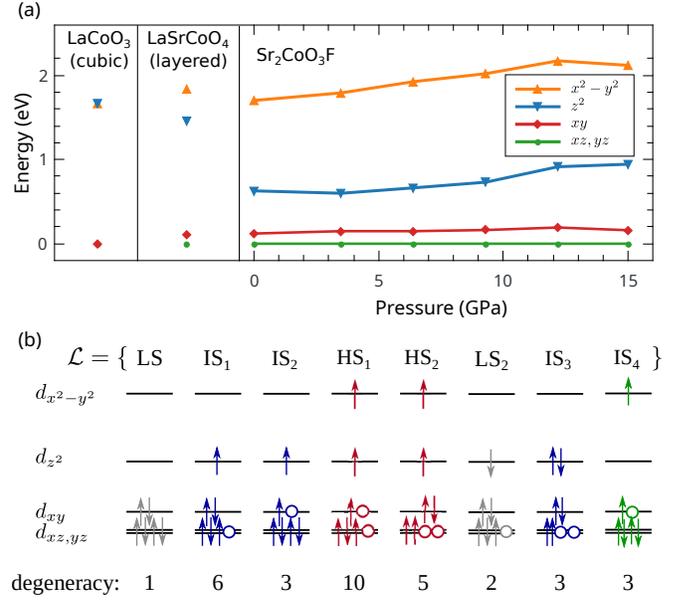}
\caption{(a) Diagonal elements $h^{ii}_{\kappa\kappa}$ of the local Hamiltonian 
as functions of pressure for \scof{}. The same quantities for other two compounds at ambient pressure are provided for comparison. (b) Spin configurations of the lowest atomic multiplets of \scof{} in the absence of SOC.}
\label{states}
\end{figure}

The tight-binding parameters for the Co-$3d$ bands are 
obtained by projecting the LDA Hamiltonian to Wannier orbital basis \cite{wien2wannier,wannier90}. 
With these, the system is described by the five-orbital Hubbard Hamiltonian
\begin{equation}\label{ham_FH}
 \hat{\cal H} = 
 \sum_{i}\hat{\cal H}^{(i)}_{\text{at}}
 +\sum_{ij}\hat{\cal H}^{(ij)}_{t},
\end{equation}
where
\begin{eqnarray}
    &&\hat{\cal H}^{(i)}_{\text{at}} =
    \sum_{\kappa\lambda}h_{\kappa\lambda}^{ii}
    \hat{c}^\dag_{i\kappa}\hat{c}^{\phantom{\dag}}_{i\lambda}
    + \sum_{\kappa\lambda\mu\nu}U_{\kappa\lambda\mu\nu}
    \hat{c}^\dag_{i\kappa}\hat{c}^\dag_{i\lambda}
    \hat{c}^{\phantom{\dag}}_{i\nu}\hat{c}^{\phantom{\dag}}_{i\mu},
    \label{ham_int}
    \\
    &&\hat{\cal H}^{(ij)}_{t}=\sum_{\kappa\lambda}
    h_{\kappa\lambda}^{ij}\hat{c}^\dag_{i\kappa}
    \hat{c}^{\phantom{\dag}}_{j\lambda},~~i\neq j.
    \label{ham_hop}
\end{eqnarray}
Here, $\hat{c}^\dagger_{{i}\kappa}$ ($\hat{c}^{\phantom{\dag}}_{{i}\kappa}$) is a fermionic creation (annihilation) operator, $i$ and $j$ refer to the site, while $\kappa,\lambda,\mu,\nu$ are the combined orbital and spin state indices.
The matrix $h_{\kappa\lambda}^{ii}$ describes the crystal-field splitting and spin-orbit coupling (SOC) \footnote{The employed amplitude of SOC $\zeta=56$~meV is based on the electronic-structure analysis in \lsco{}~\cite{afonso18}.}. The crystal-field splitting increases with pressure~$P$, as summarized in Fig.~\ref{states}(a).
To determine the local interaction $U_{\kappa\lambda\mu\nu}{(F_{0},F_{2},F_{4})}$, 
we fix $F_0=3.0$~eV~\cite{karolak15}, $F_4/F_2= 0.625$~\cite{Pavarini1,Pavarini2}, and treat the Hund's coupling~${J}=(F_2+F_4)/14$ as a tunable parameter.
%
%

\subsection{Strong coupling expansion and mean-field approximation}\label{sec.sc}
We diagonalize the local Hamiltonian~$\hat{\cal H}^{(i)}_{\text{at}}$ to obtain the set $\cal{L}$ of low-energy atomic states of the system. 
The multiplets with lowest energies are shown in Fig.~\ref{states}(b). 
An important feature of the present Slater-Condon term of the local Coulomb interaction is that IS$_1$ and HS$_1$ states have lower energies than IS$_2$ and HS$_2$, respectively.
This contrasts with simpler (density-density or Slater-Kanamori) parametrization of interactions, or  with the non-interacting limit, where IS$_2$ and HS$_2$ become lower due to the tetragonal crystal-field splitting.

Next, we apply the Schrieffer-Wolff transformation to second order in the hopping $\hat{\cal H}^{(ij)}_{t}$ and keep the nearest-neighbor terms only; see also Ref.~\cite{afonso18} for details. For each pressure~$P$, the above procedure leads to an effective Hamiltonian that spans a finite set of the lowest atomic states [see, e.g., Fig.~\ref{states}(b)]. 
Using the Schwinger boson representation, the effective Hamiltonian can be written as follows
\begin{equation}\label{Hbos}
 \hat{\cal H}_{\rm eff}=  \sum_{\bf{ i}}
 \sum_{\bf{e}=\pm a,\pm b}
 \sum_{\alpha\beta\gamma\delta \in \cal{L}}
 \varepsilon_{\alpha\beta\gamma\delta}^{(\bf{e})}
 \left(\hat{d}^\dag_{\bf{i}\alpha} \hat{d}^{\phantom{\dag}}_{\bf{i}\beta}\right)
 \left(\hat{d}^\dag_{\bf{i+e}\gamma} \hat{d}^{\phantom{\dag}}_{\bf{i+e}\delta}\right),
\end{equation}
where $\hat{d}^\dagger_{\bf{i}\alpha}$ ($\hat{d}^{\phantom{\dag}}_{\bf{i}\alpha}$) is a creation (annihilation) operator of boson representing the atomic state $\alpha$ on the site~$\bf{i}$. Hard-core constraint 
$\sum_\alpha \hat{d}^{\dag}_{i\alpha}\hat{d}^{\phantom{\dag}}_{i\alpha}=\hat{1}$ is imposed on the physical states.


Due to the complexity of the model~\eqref{Hbos} for analysis with exact methods, we apply the mean-field  decoupling in the local basis.
We restrict the SC-MF calculations to a two-site unit cell, which allows us to account for both the SSO and AFM ordering.
The access to local observables of the $\braket{\hat{d}^\dagger_{i\alpha} \hat{d}^{\phantom{\dagger}}_{i\beta}}$ type allows us to probe orders involving local superpositions of different multiplets, e.g., excitonic condensates \cite{kunes15}.

The excitation spectrum for the model~\eqref{Hbos} can be obtained by the generalized spin-wave approach~\cite{svb01,afonso18}.
With the intent to search for an incipient excitonic instability,
we apply it in the LS phase. Here, one component, $\alpha={\rm LS}$, is the vacuum state. 
The terms in Eq.~\eqref{Hbos}, which contain three or four operators with $\alpha\neq{\rm LS}$ describing interactions between excitations, are omitted. Therefore, the spin-wave analysis is restricted to the low-temperature region, where the thermal population of excited states can be neglected.

\subsection{LDA+DMFT approach}

The LDA+DMFT calculations are performed using the \textsc{w2dynamics}~\cite{w2dynamics} implementation of the strong-coupling continuous-time quantum Monte Carlo solver for the Anderson impurity model with full Coulomb vertex~$U_{\kappa\lambda\mu\nu}$. The superstate-sampling algorithm is used to boost the efficiency of the calculations~\cite{Kowalski18}. The small off-diagonal elements in the hybridization function due to the distortion of the Co atom are neglected. After the DMFT self-consistent calculation is converged, the spectral functions along the real-frequency axis are obtained by the analytic continuation using the maximum entropy method~\cite{w2dynamics,Geffroy18}. To study SSO and AFM ordering, we employ a unit cell containing two Co atoms. The calculations are performed at $T=290$~K.

\section{Results and discussion}\label{res}

At high pressure (large crystal-field splitting), the global ground state of \scof{} corresponds to all atoms in the LS state.
The dominant effect of pressure is to change the crystal-field splitting; see also Fig.~\ref{states}(a).
As pressure decreases, the energy difference between LS and other [intermediate-spin (IS) and HS] states reduces.
When the energy of atomic excitations nears the energy of LS states, the global ground state changes.
This is the so-called ``spin state crossover'' regime, where interactions between excitations and their mobility become important.

The HS excitations interact repulsively on nearest-neighbor (NN) bonds. The reason for that is the superexchange mechanism. A HS state surrounded by LS sites can lower its energy by a number of virtual hopping processes. In contrast, when two HS states with the same orbital and spin character occupy NN sites, the virtual hopping is blocked by the Pauli principle.
The interaction is thus strongly repulsive (in \scof{}, it is calculated to be around $0.2$~eV per bond). 
For HS states with different spin/orbital character, Pauli blocking is not complete and
the NN repulsion is thus weaker. 
Note that even in AFM with antiferroorbital (AFO) arrangement, two HS states still interact repulsively on NN bonds (typical values are around 30~meV per bond). The interactions involving IS states have more diverse structure ranging from weakly repulsive to weakly attractive depending on the spin, orbital and bond orientations.

With  decreasing pressure, the energy gap between LS and exited states shrinks.
If the lowest excitations are localized but strongly interact on NN bonds, a checkerboard arrangement of these and LS states is expected \cite{kunes11,karolak15}.
However, as soon as some low-lying excitations are mobile, a competing excitonic condensation (EC) instability becomes important~\cite{kunes15,sotnikov16,afonso18}.

\begin{figure}
\includegraphics[width=\columnwidth]{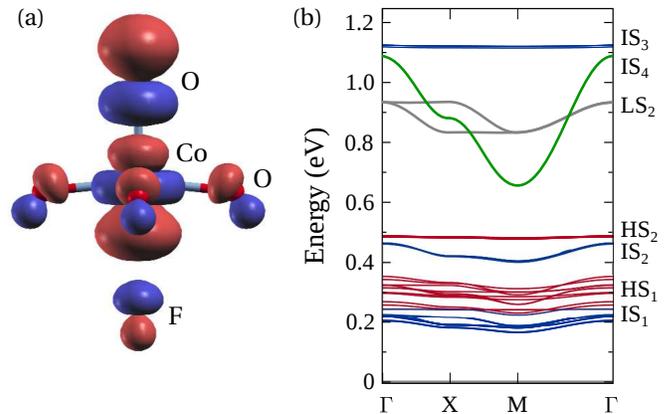}
\caption{Co-$d_{z^2}$ Wannier function from the $d$-only model with the corresponding weights on adjacent atoms (``Wannier tails'') plotted as isosurfaces colored by the sign (a). Energies of the excitations (b) obtained from the SC expansion with the spin-wave analysis at $P=12.2$~GPa, $J=0.62$~eV, and $\zeta=56$~meV.}
\label{disp}
\end{figure}

We start our presentation in the high-pressure LS phase.
To examine the EC scenario in \scof{}, we employed the spin-wave analysis.
It reveals flat bands for the lowest excitations; see Fig.~\ref{disp}(b).
%
%
Only the IS$_4$ excitations have a substantial dispersion. 
The flat dispersion of the IS$_1$ and IS$_2$ excitations originates from small $d_{z^2}$--$d_{z^2}$ hopping amplitude between the nearest neighbors. For comparison, the $d_{z^2}$--$d_{z^2}$ hopping amplitude is 147~meV~\cite{wang18prb} and 133~meV~\cite{afonso18} in \laco{} and \lsco{}, respectively, while in \scof{} it averages 14~meV in the studied range of the applied pressure, i.e., an order of magnitude smaller. This can be understood by
considering the shape of the $d_{z^2}$ Wannier orbitals shown in Fig.~\ref{disp}(a).
Asymmetric hybridization with the apical O and F atoms causes tilting of the Wannier tails on the in-plane O sites. As a result, the overlapping tails of the neighboring Wannier $d_{z^2}$ orbitals are almost orthogonal.
In contrast, the $d_{x^2-y^2}\otimes d_{xy}$ IS$_4$ excitons possess
a dispersion with a large band width, similar to LaCoO$_3$~\cite{wang18prb}. 
Nevertheless, the large crystal-field energy
of the $d_{x^2-y^2}$ state gives rise to a 0.7-eV gap, which excludes any
excitonic instability.

\begin{figure}
\includegraphics[width=\columnwidth]{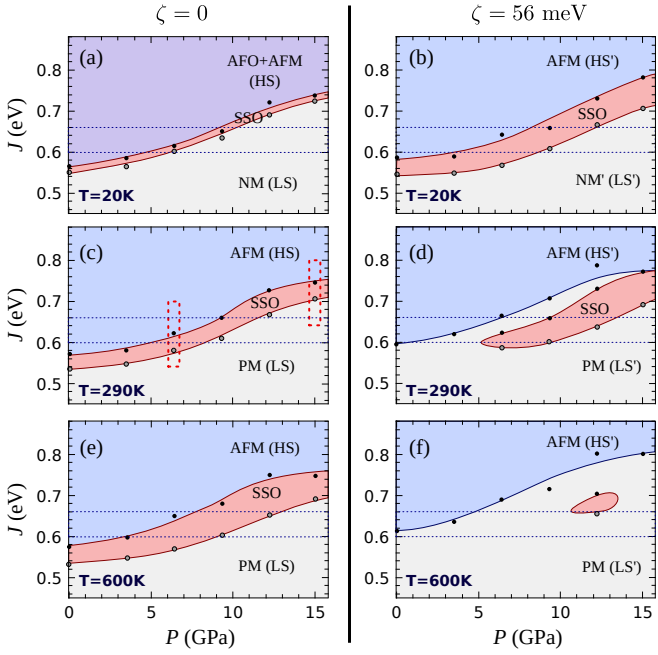}
\caption{Phase diagrams obtained with SC-MF in the absence of SOC (left) and with $\zeta=56$~meV (right) at different temperatures. The horizontal lines correspond to the upper and lower bounds for $J$ to support the experimental observations.
In (c), the regions at $P=6.4$~GPa  and $P=15$~GPa are indicated, where a quantitative comparison with LDA+DMFT was performed.}
\label{pdiags}
\end{figure}

We use the SC-MF analysis to study phase diagram of \scof{} under pressure, shown in Fig.~\ref{pdiags}.
%
%
There are three phases: 
(i) nonmagnetic (NM) or thermally induced paramagnetic (PM) LS phase;
(ii) paramagnetic SSO phase;
(iii) AFM HS (with an admixture of IS) phase.
SOC only slightly affects the overall structure of phase diagrams. At low temperatures, see Figs.~\ref{pdiags}(a) and \ref{pdiags}(b), SOC removes additional antiferro-orbital (AFO) ordering of the AFM (HS) phase. It also stabilizes SSO at low temperatures leading to wider regions in the $P$-$J$ planes. With the temperature increase, SSO order becomes more susceptible to thermal fluctuations in the presence of SOC, see Figs.~\ref{pdiags}(c)--\ref{pdiags}(f).

\begin{figure}
\includegraphics[width=\columnwidth]{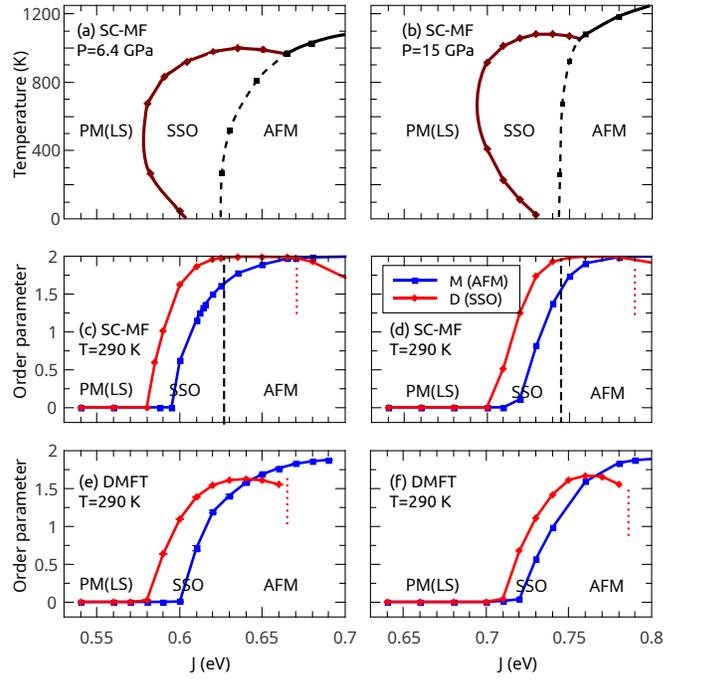}
\caption{$J$-$T$ phase diagrams (upper row) and comparison of two theoretical approaches at pressures $P=6.4$~GPa (left column) and $P=15$~GPa (right column), $T=290$~K (c)--(f), and $\zeta=0$. The vertical red dotted lines indicate the upper bound for the stability of SSO in the region of coexistence of SSO and AFM. The black dashed lines correspond to positions of the first-order SSO-AFM transitions obtained from the comparison of the mean-field free energies.}
\label{comparison}
\end{figure}

A noticeable broadening of the SSO phase with temperature in Figs.~\ref{pdiags}(a), \ref{pdiags}(c), and \ref{pdiags}(e), as well as the re-entrant PM-SSO-PM features in Fig.~\ref{pdiags}(d) at fixed $J\in[0.6,0.75]$~eV, resemble the Blume-Emery-Griffiths (BEG) model \cite{BEG}. 
Indeed, the $J$-$T$ phase diagrams in Figs.~\ref{comparison}(a) and \ref{comparison}(b) have the same shape as those
of the BEG model~\cite{hoston91,kunes15}, where the Hund's coupling $J$ controls the splitting
between the LS and HS states, i.e., it plays the same role as the single-ion anisotropy term in the BEG model. The shrinking of the area occupied by the SSO phase both at low and high temperatures arises from the dual role played by temperature. On one hand, temperature generates the HS sites necessary for the SSO phase to form. On the other hand, thermal fluctuations destroy the SSO order when the temperature is too high.
Note that the PM-SSO and PM-AFM transitions are continuous, while the SSO-AFM transition is of the first order, both in \scof{} and in the BEG model.

The broadening of the SSO phase in the presence of SOC at low temperature can be attributed to the splitting of the HS and IS energy multiplets.
The lower critical temperatures for both SSO and AFM phases in the presence of SOC are due to renormalization of the coupling constants in the effective model~\eqref{Hbos}.
%
%

To confirm reliability of the SC-MF results, we compare local observables from two approaches, SC-MF and DMFT, in two parameter regimes [marked regions in Fig.~\ref{pdiags}(c)] in Figs.~\ref{comparison}(c)--\ref{comparison}(f).
The staggered AFM and SSO order parameters are defined as follows:
\begin{eqnarray}
  &&M=\sum_{i=1,2}(-1)^i\sum_{m=1}^5 (n_{i,m\uparrow}-n_{i,m\downarrow}),
  \\
  &&D=\sum_{i=1,2}(-1)^i[n_{i,{\rm HS_1}}-(n_{i,{\rm LS}} + n_{i,{\rm IS_1}})],
\end{eqnarray}
where $i$ and $m$ are the sublattice and orbital indices, respectively~\footnote{In DMFT, the SSO order parameter~$D$ is estimated approximately from the electron densities on different $e_g$ and $t_{2g}$ orbitals with restricting to the subspace of three lowest multiplets LS, IS$_1$, and HS$_1$.}, and
$n^{\phantom{\dagger}}_{i\alpha}\equiv\langle \hat{d}^\dagger_{i,\alpha}\hat{d}^{\phantom{\dagger}}_{i\alpha} \rangle$.

Our results show that the SC-MF approach can be viewed as a sufficient method to estimate positions of the phase boundaries for the system under study. At the same time, as expected, it overestimates the magnitudes of the order parameters in the symmetry-broken phases (and, most likely, critical temperatures). Nevertheless, according to additional analysis, it is more precise in comparison with the commonly used restriction to the density-density type of interactions. In addition, SC-MF allows us to analyze relevant physical observables in a wide temperature range and include the effect of SOC in a straightforward manner.

\begin{figure}
\includegraphics[width=\columnwidth]{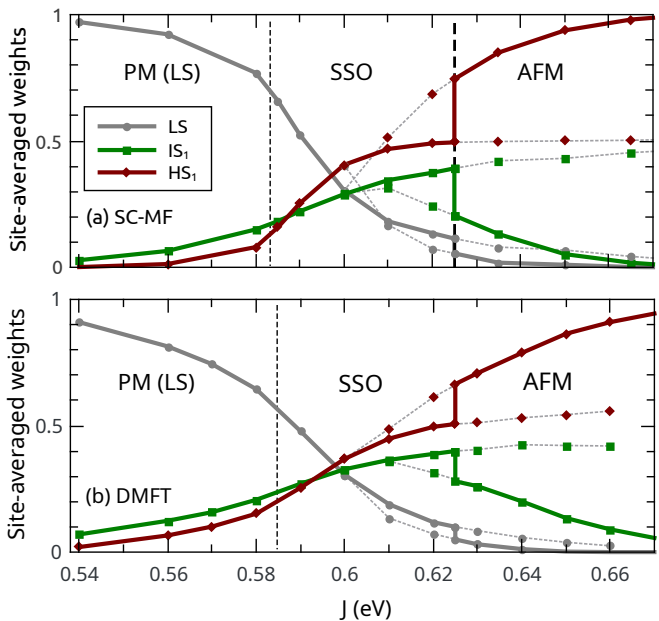}
\caption{ Atomic weights obtained by SC-MF (a) and DMFT (b) at $P=6.4$~GPa, $\zeta=0$, and $T=290$~K. Metastable solutions are indicated by dotted lines. In (b), the position of the first-order transition is taken the same as in SC-MF.}
\label{weights}
\end{figure}

\begin{figure}
\includegraphics[width=\columnwidth]{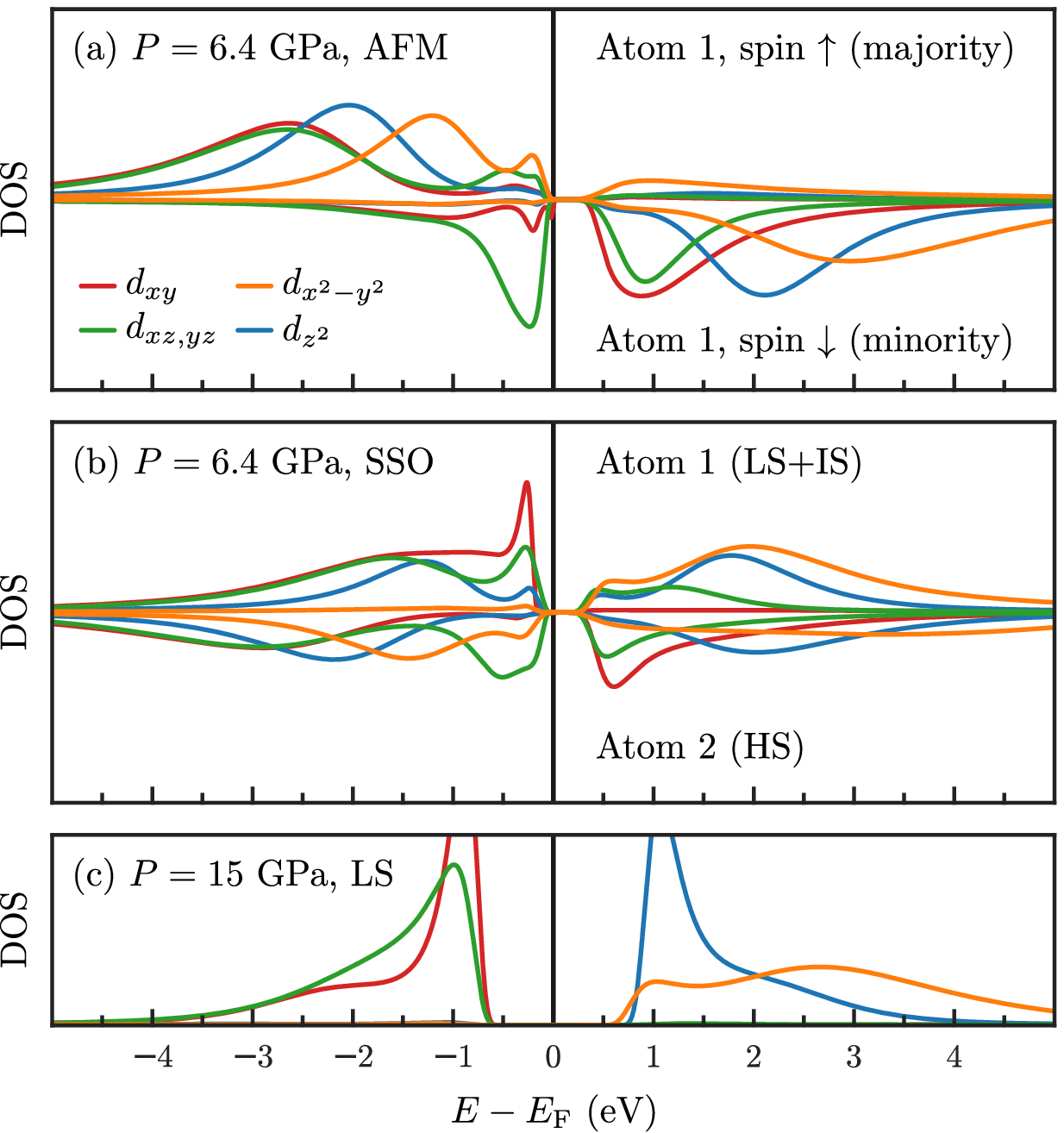}
\includegraphics[width=\columnwidth]{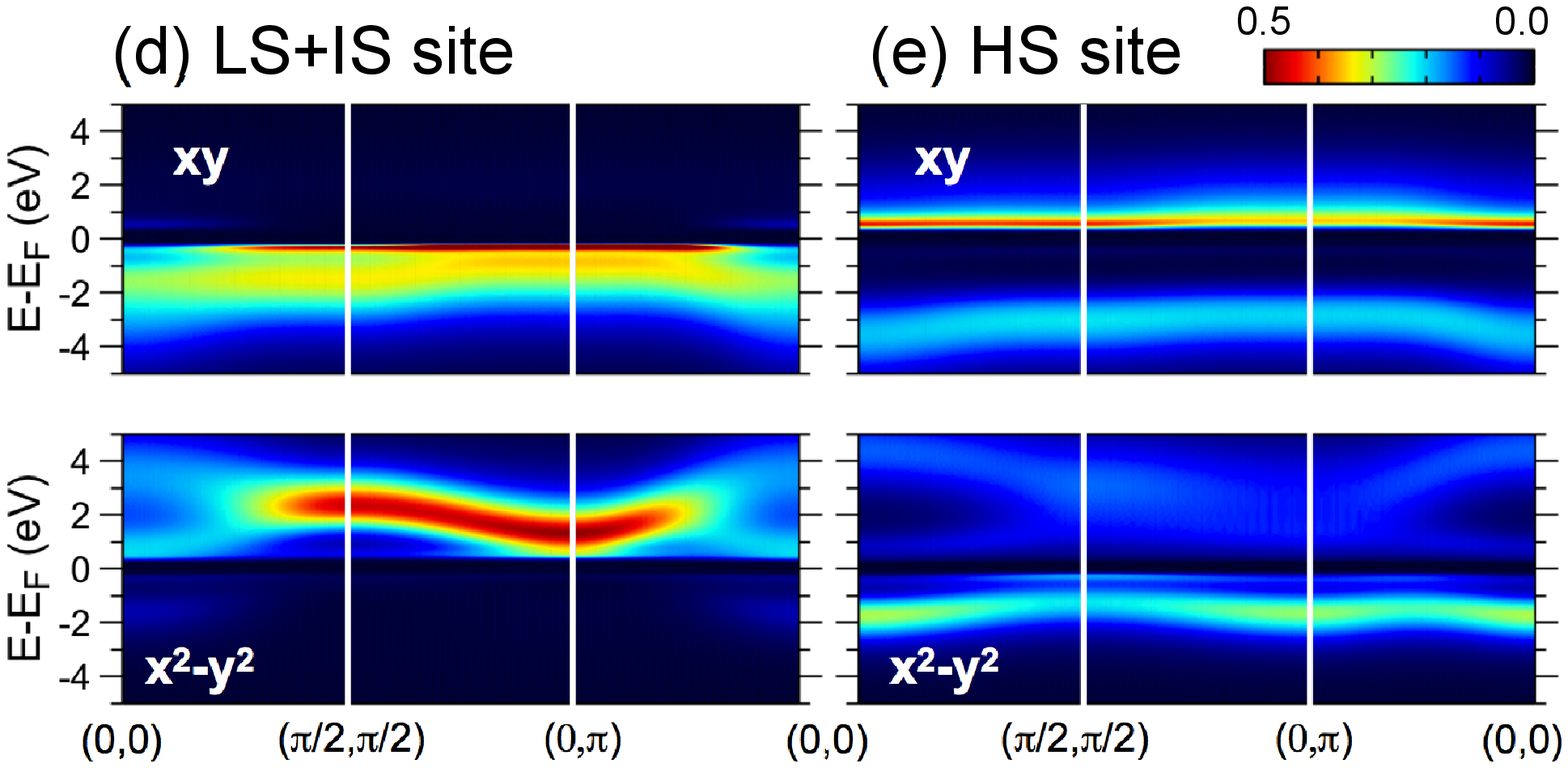}
\caption{Spectral functions obtained in the AFM (a) and SSO (b) phases at $P=6.4$~GPa and in the LS phase (c) at $P=15$~GPa and $T=290$~K. In panels (d) and (e) we show the $k$-resolved spectral functions of $d_{xy}$ and $d_{x^2-y^2}$ character in the SSO phase.}
\label{spectra}
\end{figure}

Although the main driving mechanism for SSO is the strong NN repulsion between HS states on the lattice, there is a sizable contribution of IS states in this compound. In Fig.~\ref{weights} we show the atomic weights corresponding to the converged solutions from two approaches.
%
%
In the SSO phase, the weight of IS states increases with $J$ due to substitution of LS by IS, so that close to the SSO-AFM transition the state becomes rather of the HS-IS than of the HS-LS type.
Using the maximum-entropy method to analytically continue the DMFT
data, we can compare one-particle spectral functions of \scof{} of different phases in Fig.~\ref{spectra}. The spectra were obtained
at two pressures, $P=6.4$~GPa and $P=15$~GPa, with the Hund's exchange $J=0.64$~eV, which matches best the experimental observations.
%
The estimated gaps for the AFM, SSO, and LS phases are, respectively, 0.25, 0.21, and 1.35 eV.    
The spectral weights above the Fermi energy~$E_{\rm F}$ of the $d_{x^2-y^2}$ and $d_{xz/yz}$ states in the AFM (spin $\uparrow$) and SSO (atom 1) panels, respectively, confirm significant contributions of IS$_1$ states that are also apparent in Fig.~\ref{weights}. In Figs.~\ref{spectra}(d) and \ref{spectra}(e) we show the $k$-resolved spectral functions
of the $d_{x^2-y^2}$ and $d_{xy}$ orbitals. The spectral functions for both orbitals reflect the doubling of the unit cell and formation of Hubbard bands on the HS site.
Since photoemission measurements are not possible under pressure, the experimental observation of the SSO phase is likely to rely on local probes that are able to detect formation of two distinct Co sites such as Raman or M\"ossbauer spectroscopy or nuclear magnetic resonance.

\section{Conclusions}\label{con}
We have reported DFT+DMFT and DFT+SC-MF study of the layered cobaltite \scof{} under pressure. Unlike in the
isoelectronic LaCoO$_3$ and LaSrCoO$_4$, mobile IS excitations were found to play no role in low-energy physics of \scof{}. Mobile $d_{x^2-y^2}\otimes d_{xy}$ IS$_4$ excitations exhibit a gap over 0.5~eV due to tetragonal  crystal-field splitting, while the other IS excitations involving $d_{z^2}$ orbital have substantially reduced mobility due to asymmetric hybridization between Co and apical O and F atoms. 
The low-energy physics of \scof{} is governed by NN interactions between atom-bound multiplet excitations and our SC model reduces to a generalized BEG model. As in the BEG model~\cite{BEG,hoston91}, we find the SSO phase between LS and AFM phases. The SC-MF results are confirmed by matching DMFT results. 

The existence of sublattices with two distinct atomic states of Co in the SSO state renders its detection straightforward. For example, a sizable Co-O bond-length disproportionation~\cite{knizek09} is expected. 
The SSO state has been invoked in the context of LaCoO$_3$~\cite{Bari1972,kunes11,karolak15}, but never  experimentally observed. We suppose mobile low-energy IS excitations~\cite{sotnikov16,wang18prb} to be the reason. The absence or presence of these in \scof{} and LaCoO$_3$, respectively, can
explain their different magnetic properties --- the AFM ordering of \scof{} {vs} ferromagnetic correlations in bulk LaCoO$_3$~\cite{tomiyasu18pre} and ferromagnetic ordering in strained LaCoO$_3$ films~\cite{fuchs07,freeland08}.
\scof{} thus provides an important reference system to understand the physics of Co$^{3+}$ perovskites.

\section{Acknowledgments}
The authors thank M.~Pickem, J.~Kaufmann, and A.~Hausoel for technical assistance with the \textsc{w2dynamics} package.
This work has received funding from the European Research Council (ERC) under the European Union's Horizon 2020 research
and innovation programme (Grant Agreement No. 646807-EXMAG).
Access to computing and storage facilities provided by the Vienna Scientific Cluster (VSC) is greatly appreciated.

\bibliography{SSOpressure}	

\begin{thebibliography}{46}%
\makeatletter
\providecommand \@ifxundefined [1]{%
 \@ifx{#1\undefined}
}%
\providecommand \@ifnum [1]{%
 \ifnum #1\expandafter \@firstoftwo
 \else \expandafter \@secondoftwo
 \fi
}%
\providecommand \@ifx [1]{%
 \ifx #1\expandafter \@firstoftwo
 \else \expandafter \@secondoftwo
 \fi
}%
\providecommand \natexlab [1]{#1}%
\providecommand \enquote  [1]{``#1''}%
\providecommand \bibnamefont  [1]{#1}%
\providecommand \bibfnamefont [1]{#1}%
\providecommand \citenamefont [1]{#1}%
\providecommand \href@noop [0]{\@secondoftwo}%
\providecommand \href [0]{\begingroup \@sanitize@url \@href}%
\providecommand \@href[1]{\@@startlink{#1}\@@href}%
\providecommand \@@href[1]{\endgroup#1\@@endlink}%
\providecommand \@sanitize@url [0]{\catcode `\\12\catcode `\$12\catcode
  `\&12\catcode `\#12\catcode `\^12\catcode `\_12\catcode `\%12\relax}%
\providecommand \@@startlink[1]{}%
\providecommand \@@endlink[0]{}%
\providecommand \url  [0]{\begingroup\@sanitize@url \@url }%
\providecommand \@url [1]{\endgroup\@href {#1}{\urlprefix }}%
\providecommand \urlprefix  [0]{URL }%
\providecommand \Eprint [0]{\href }%
\providecommand \doibase [0]{http://dx.doi.org/}%
\providecommand \selectlanguage [0]{\@gobble}%
\providecommand \bibinfo  [0]{\@secondoftwo}%
\providecommand \bibfield  [0]{\@secondoftwo}%
\providecommand \translation [1]{[#1]}%
\providecommand \BibitemOpen [0]{}%
\providecommand \bibitemStop [0]{}%
\providecommand \bibitemNoStop [0]{.\EOS\space}%
\providecommand \EOS [0]{\spacefactor3000\relax}%
\providecommand \BibitemShut  [1]{\csname bibitem#1\endcsname}%
\let\auto@bib@innerbib\@empty
\bibitem [{\citenamefont {Tanabe}\ and\ \citenamefont
  {Sugano}(1954)}]{Tanabe1954}%
  \BibitemOpen
  \bibfield  {author} {\bibinfo {author} {\bibfnamefont {Y.}~\bibnamefont
  {Tanabe}}\ and\ \bibinfo {author} {\bibfnamefont {S.}~\bibnamefont
  {Sugano}},\ }\href {\doibase 10.1143/JPSJ.9.753} {\bibfield  {journal}
  {\bibinfo  {journal} {J. Phys. Soc. Jpn.}\ }\textbf {\bibinfo {volume} {9}},\
  \bibinfo {pages} {753} (\bibinfo {year} {1954})}\BibitemShut {NoStop}%
\bibitem [{\citenamefont {Goodenough}(1971)}]{Goodenough71}%
  \BibitemOpen
  \bibfield  {author} {\bibinfo {author} {\bibfnamefont {J.~B.}\ \bibnamefont
  {Goodenough}},\ }in\ \href {\doibase 10.1016/S0081-1947(08)60740-7} {\emph
  {\bibinfo {booktitle} {Progress in Solid State Chemistry}}},\ Vol.~\bibinfo
  {volume} {5},\ \bibinfo {editor} {edited by\ \bibinfo {editor} {\bibfnamefont
  {H.}~\bibnamefont {Reiss}}}\ (\bibinfo  {publisher} {Pergamon, Oxford},\
  \bibinfo {year} {1971})\ pp.\ \bibinfo {pages} {145--399}\BibitemShut
  {NoStop}%
\bibitem [{\citenamefont {Wu}\ and\ \citenamefont {Leighton}(2003)}]{wu03}%
  \BibitemOpen
  \bibfield  {author} {\bibinfo {author} {\bibfnamefont {J.}~\bibnamefont
  {Wu}}\ and\ \bibinfo {author} {\bibfnamefont {C.}~\bibnamefont {Leighton}},\
  }\href {\doibase 10.1103/PhysRevB.67.174408} {\bibfield  {journal} {\bibinfo
  {journal} {Phys. Rev. B}\ }\textbf {\bibinfo {volume} {67}},\ \bibinfo
  {pages} {174408} (\bibinfo {year} {2003})}\BibitemShut {NoStop}%
\bibitem [{\citenamefont {Phelan}\ \emph {et~al.}(2006)\citenamefont {Phelan},
  \citenamefont {Louca}, \citenamefont {Rosenkranz}, \citenamefont {Lee},
  \citenamefont {Qiu}, \citenamefont {Chupas}, \citenamefont {Osborn},
  \citenamefont {Zheng}, \citenamefont {Mitchell}, \citenamefont {Copley},
  \citenamefont {Sarrao},\ and\ \citenamefont {Moritomo}}]{Phelan2006}%
  \BibitemOpen
  \bibfield  {author} {\bibinfo {author} {\bibfnamefont {D.}~\bibnamefont
  {Phelan}}, \bibinfo {author} {\bibfnamefont {D.}~\bibnamefont {Louca}},
  \bibinfo {author} {\bibfnamefont {S.}~\bibnamefont {Rosenkranz}}, \bibinfo
  {author} {\bibfnamefont {S.-H.}\ \bibnamefont {Lee}}, \bibinfo {author}
  {\bibfnamefont {Y.}~\bibnamefont {Qiu}}, \bibinfo {author} {\bibfnamefont
  {P.~J.}\ \bibnamefont {Chupas}}, \bibinfo {author} {\bibfnamefont
  {R.}~\bibnamefont {Osborn}}, \bibinfo {author} {\bibfnamefont
  {H.}~\bibnamefont {Zheng}}, \bibinfo {author} {\bibfnamefont {J.~F.}\
  \bibnamefont {Mitchell}}, \bibinfo {author} {\bibfnamefont {J.~R.~D.}\
  \bibnamefont {Copley}}, \bibinfo {author} {\bibfnamefont {J.~L.}\
  \bibnamefont {Sarrao}}, \ and\ \bibinfo {author} {\bibfnamefont
  {Y.}~\bibnamefont {Moritomo}},\ }\href {\doibase
  10.1103/PhysRevLett.96.027201} {\bibfield  {journal} {\bibinfo  {journal}
  {Phys. Rev. Lett.}\ }\textbf {\bibinfo {volume} {96}},\ \bibinfo {pages}
  {027201} (\bibinfo {year} {2006})}\BibitemShut {NoStop}%
\bibitem [{\citenamefont {Fuchs}\ \emph {et~al.}(2007)\citenamefont {Fuchs},
  \citenamefont {Pinta}, \citenamefont {Schwarz}, \citenamefont {Schweiss},
  \citenamefont {Nagel}, \citenamefont {Schuppler}, \citenamefont {Schneider},
  \citenamefont {Merz}, \citenamefont {Roth},\ and\ \citenamefont
  {v.~L\"ohneysen}}]{fuchs07}%
  \BibitemOpen
  \bibfield  {author} {\bibinfo {author} {\bibfnamefont {D.}~\bibnamefont
  {Fuchs}}, \bibinfo {author} {\bibfnamefont {C.}~\bibnamefont {Pinta}},
  \bibinfo {author} {\bibfnamefont {T.}~\bibnamefont {Schwarz}}, \bibinfo
  {author} {\bibfnamefont {P.}~\bibnamefont {Schweiss}}, \bibinfo {author}
  {\bibfnamefont {P.}~\bibnamefont {Nagel}}, \bibinfo {author} {\bibfnamefont
  {S.}~\bibnamefont {Schuppler}}, \bibinfo {author} {\bibfnamefont
  {R.}~\bibnamefont {Schneider}}, \bibinfo {author} {\bibfnamefont
  {M.}~\bibnamefont {Merz}}, \bibinfo {author} {\bibfnamefont {G.}~\bibnamefont
  {Roth}}, \ and\ \bibinfo {author} {\bibfnamefont {H.}~\bibnamefont
  {v.~L\"ohneysen}},\ }\href {\doibase 10.1103/PhysRevB.75.144402} {\bibfield
  {journal} {\bibinfo  {journal} {Phys. Rev. B}\ }\textbf {\bibinfo {volume}
  {75}},\ \bibinfo {pages} {144402} (\bibinfo {year} {2007})}\BibitemShut
  {NoStop}%
\bibitem [{\citenamefont {Fujioka}\ \emph {et~al.}(2013)\citenamefont
  {Fujioka}, \citenamefont {Yamasaki}, \citenamefont {Nakao}, \citenamefont
  {Kumai}, \citenamefont {Murakami}, \citenamefont {Nakamura}, \citenamefont
  {Kawasaki},\ and\ \citenamefont {Tokura}}]{fujioka13prl}%
  \BibitemOpen
  \bibfield  {author} {\bibinfo {author} {\bibfnamefont {J.}~\bibnamefont
  {Fujioka}}, \bibinfo {author} {\bibfnamefont {Y.}~\bibnamefont {Yamasaki}},
  \bibinfo {author} {\bibfnamefont {H.}~\bibnamefont {Nakao}}, \bibinfo
  {author} {\bibfnamefont {R.}~\bibnamefont {Kumai}}, \bibinfo {author}
  {\bibfnamefont {Y.}~\bibnamefont {Murakami}}, \bibinfo {author}
  {\bibfnamefont {M.}~\bibnamefont {Nakamura}}, \bibinfo {author}
  {\bibfnamefont {M.}~\bibnamefont {Kawasaki}}, \ and\ \bibinfo {author}
  {\bibfnamefont {Y.}~\bibnamefont {Tokura}},\ }\href {\doibase
  10.1103/PhysRevLett.111.027206} {\bibfield  {journal} {\bibinfo  {journal}
  {Phys. Rev. Lett.}\ }\textbf {\bibinfo {volume} {111}},\ \bibinfo {pages}
  {027206} (\bibinfo {year} {2013})}\BibitemShut {NoStop}%
\bibitem [{\citenamefont {Guo}\ \emph {et~al.}(2016)\citenamefont {Guo},
  \citenamefont {Hu}, \citenamefont {Pi}, \citenamefont {Tjeng},\ and\
  \citenamefont {Komarek}}]{guo16}%
  \BibitemOpen
  \bibfield  {author} {\bibinfo {author} {\bibfnamefont {H.}~\bibnamefont
  {Guo}}, \bibinfo {author} {\bibfnamefont {Z.}~\bibnamefont {Hu}}, \bibinfo
  {author} {\bibfnamefont {T.-W.}\ \bibnamefont {Pi}}, \bibinfo {author}
  {\bibfnamefont {L.~H.}\ \bibnamefont {Tjeng}}, \ and\ \bibinfo {author}
  {\bibfnamefont {A.~C.}\ \bibnamefont {Komarek}},\ }\href {\doibase
  10.3390/cryst6080098} {\bibfield  {journal} {\bibinfo  {journal} {Crystals}\
  }\textbf {\bibinfo {volume} {6}},\ \bibinfo {pages} {98} (\bibinfo {year}
  {2016})}\BibitemShut {NoStop}%
\bibitem [{\citenamefont {Tsubouchi}\ \emph {et~al.}(2004)\citenamefont
  {Tsubouchi}, \citenamefont {Ky\^omen}, \citenamefont {Itoh},\ and\
  \citenamefont {Oguni}}]{tsubouchi04}%
  \BibitemOpen
  \bibfield  {author} {\bibinfo {author} {\bibfnamefont {S.}~\bibnamefont
  {Tsubouchi}}, \bibinfo {author} {\bibfnamefont {T.}~\bibnamefont {Ky\^omen}},
  \bibinfo {author} {\bibfnamefont {M.}~\bibnamefont {Itoh}}, \ and\ \bibinfo
  {author} {\bibfnamefont {M.}~\bibnamefont {Oguni}},\ }\href {\doibase
  10.1103/PhysRevB.69.144406} {\bibfield  {journal} {\bibinfo  {journal} {Phys.
  Rev. B}\ }\textbf {\bibinfo {volume} {69}},\ \bibinfo {pages} {144406}
  (\bibinfo {year} {2004})}\BibitemShut {NoStop}%
\bibitem [{\citenamefont {Hejtm\'anek}\ \emph {et~al.}(2010)\citenamefont
  {Hejtm\'anek}, \citenamefont {\ifmmode~\check{S}\else \v{S}\fi{}antav\'a},
  \citenamefont {Kn\'{\i}\ifmmode~\check{z}\else \v{z}\fi{}ek}, \citenamefont
  {Mary\ifmmode~\check{s}\else \v{s}\fi{}ko}, \citenamefont {Jir\'ak},
  \citenamefont {Naito}, \citenamefont {Sasaki},\ and\ \citenamefont
  {Fujishiro}}]{hejtmanek10}%
  \BibitemOpen
  \bibfield  {author} {\bibinfo {author} {\bibfnamefont {J.}~\bibnamefont
  {Hejtm\'anek}}, \bibinfo {author} {\bibfnamefont {E.}~\bibnamefont
  {\ifmmode~\check{S}\else \v{S}\fi{}antav\'a}}, \bibinfo {author}
  {\bibfnamefont {K.}~\bibnamefont {Kn\'{\i}\ifmmode~\check{z}\else
  \v{z}\fi{}ek}}, \bibinfo {author} {\bibfnamefont {M.}~\bibnamefont
  {Mary\ifmmode~\check{s}\else \v{s}\fi{}ko}}, \bibinfo {author} {\bibfnamefont
  {Z.}~\bibnamefont {Jir\'ak}}, \bibinfo {author} {\bibfnamefont
  {T.}~\bibnamefont {Naito}}, \bibinfo {author} {\bibfnamefont
  {H.}~\bibnamefont {Sasaki}}, \ and\ \bibinfo {author} {\bibfnamefont
  {H.}~\bibnamefont {Fujishiro}},\ }\href {\doibase 10.1103/PhysRevB.82.165107}
  {\bibfield  {journal} {\bibinfo  {journal} {Phys. Rev. B}\ }\textbf {\bibinfo
  {volume} {82}},\ \bibinfo {pages} {165107} (\bibinfo {year}
  {2010})}\BibitemShut {NoStop}%
\bibitem [{\citenamefont {Wang}\ \emph {et~al.}(2018)\citenamefont {Wang},
  \citenamefont {Hariki}, \citenamefont {Sotnikov}, \citenamefont {Frati},
  \citenamefont {Okamoto}, \citenamefont {Huang}, \citenamefont {Singh},
  \citenamefont {Huang}, \citenamefont {Tomiyasu}, \citenamefont {Du},
  \citenamefont {Kune\ifmmode~\check{s}\else \v{s}\fi{}},\ and\ \citenamefont
  {de~Groot}}]{wang18prb}%
  \BibitemOpen
  \bibfield  {author} {\bibinfo {author} {\bibfnamefont {R.-P.}\ \bibnamefont
  {Wang}}, \bibinfo {author} {\bibfnamefont {A.}~\bibnamefont {Hariki}},
  \bibinfo {author} {\bibfnamefont {A.}~\bibnamefont {Sotnikov}}, \bibinfo
  {author} {\bibfnamefont {F.}~\bibnamefont {Frati}}, \bibinfo {author}
  {\bibfnamefont {J.}~\bibnamefont {Okamoto}}, \bibinfo {author} {\bibfnamefont
  {H.-Y.}\ \bibnamefont {Huang}}, \bibinfo {author} {\bibfnamefont
  {A.}~\bibnamefont {Singh}}, \bibinfo {author} {\bibfnamefont {D.-J.}\
  \bibnamefont {Huang}}, \bibinfo {author} {\bibfnamefont {K.}~\bibnamefont
  {Tomiyasu}}, \bibinfo {author} {\bibfnamefont {C.-H.}\ \bibnamefont {Du}},
  \bibinfo {author} {\bibfnamefont {J.}~\bibnamefont
  {Kune\ifmmode~\check{s}\else \v{s}\fi{}}}, \ and\ \bibinfo {author}
  {\bibfnamefont {F.~M.~F.}\ \bibnamefont {de~Groot}},\ }\href {\doibase
  10.1103/PhysRevB.98.035149} {\bibfield  {journal} {\bibinfo  {journal} {Phys.
  Rev. B}\ }\textbf {\bibinfo {volume} {98}},\ \bibinfo {pages} {035149}
  (\bibinfo {year} {2018})}\BibitemShut {NoStop}%
\bibitem [{\citenamefont {Kune\ifmmode~\check{s}\else \v{s}\fi{}}\ and\
  \citenamefont {Augustinsk\'y}(2014)}]{kunes14a}%
  \BibitemOpen
  \bibfield  {author} {\bibinfo {author} {\bibfnamefont {J.}~\bibnamefont
  {Kune\ifmmode~\check{s}\else \v{s}\fi{}}}\ and\ \bibinfo {author}
  {\bibfnamefont {P.}~\bibnamefont {Augustinsk\'y}},\ }\href {\doibase
  10.1103/PhysRevB.89.115134} {\bibfield  {journal} {\bibinfo  {journal} {Phys.
  Rev. B}\ }\textbf {\bibinfo {volume} {89}},\ \bibinfo {pages} {115134}
  (\bibinfo {year} {2014})}\BibitemShut {NoStop}%
\bibitem [{\citenamefont {Hoshino}\ and\ \citenamefont
  {Werner}(2016)}]{hoshino16}%
  \BibitemOpen
  \bibfield  {author} {\bibinfo {author} {\bibfnamefont {S.}~\bibnamefont
  {Hoshino}}\ and\ \bibinfo {author} {\bibfnamefont {P.}~\bibnamefont
  {Werner}},\ }\href {\doibase 10.1103/PhysRevB.93.155161} {\bibfield
  {journal} {\bibinfo  {journal} {Phys. Rev. B}\ }\textbf {\bibinfo {volume}
  {93}},\ \bibinfo {pages} {155161} (\bibinfo {year} {2016})}\BibitemShut
  {NoStop}%
\bibitem [{\citenamefont {Nasu}\ \emph {et~al.}(2016)\citenamefont {Nasu},
  \citenamefont {Watanabe}, \citenamefont {Naka},\ and\ \citenamefont
  {Ishihara}}]{nasu16}%
  \BibitemOpen
  \bibfield  {author} {\bibinfo {author} {\bibfnamefont {J.}~\bibnamefont
  {Nasu}}, \bibinfo {author} {\bibfnamefont {T.}~\bibnamefont {Watanabe}},
  \bibinfo {author} {\bibfnamefont {M.}~\bibnamefont {Naka}}, \ and\ \bibinfo
  {author} {\bibfnamefont {S.}~\bibnamefont {Ishihara}},\ }\href {\doibase
  10.1103/PhysRevB.93.205136} {\bibfield  {journal} {\bibinfo  {journal} {Phys.
  Rev. B}\ }\textbf {\bibinfo {volume} {93}},\ \bibinfo {pages} {205136}
  (\bibinfo {year} {2016})}\BibitemShut {NoStop}%
\bibitem [{\citenamefont {Afonso}\ and\ \citenamefont
  {Kune\ifmmode~\check{s}\else \v{s}\fi{}}(2017)}]{afonso17}%
  \BibitemOpen
  \bibfield  {author} {\bibinfo {author} {\bibfnamefont {J.~F.}\ \bibnamefont
  {Afonso}}\ and\ \bibinfo {author} {\bibfnamefont {J.}~\bibnamefont
  {Kune\ifmmode~\check{s}\else \v{s}\fi{}}},\ }\href {\doibase
  10.1103/PhysRevB.95.115131} {\bibfield  {journal} {\bibinfo  {journal} {Phys.
  Rev. B}\ }\textbf {\bibinfo {volume} {95}},\ \bibinfo {pages} {115131}
  (\bibinfo {year} {2017})}\BibitemShut {NoStop}%
\bibitem [{\citenamefont {Hsu}\ \emph {et~al.}(2012)\citenamefont {Hsu},
  \citenamefont {Blaha},\ and\ \citenamefont {Wentzcovitch}}]{Hsu2012}%
  \BibitemOpen
  \bibfield  {author} {\bibinfo {author} {\bibfnamefont {H.}~\bibnamefont
  {Hsu}}, \bibinfo {author} {\bibfnamefont {P.}~\bibnamefont {Blaha}}, \ and\
  \bibinfo {author} {\bibfnamefont {R.~M.}\ \bibnamefont {Wentzcovitch}},\
  }\href {\doibase 10.1103/PhysRevB.85.140404} {\bibfield  {journal} {\bibinfo
  {journal} {Phys. Rev. B}\ }\textbf {\bibinfo {volume} {85}},\ \bibinfo
  {pages} {140404} (\bibinfo {year} {2012})}\BibitemShut {NoStop}%
\bibitem [{\citenamefont {Yamaguchi}\ \emph {et~al.}(2017)\citenamefont
  {Yamaguchi}, \citenamefont {Sugimoto},\ and\ \citenamefont
  {Ohta}}]{yamaguchi17}%
  \BibitemOpen
  \bibfield  {author} {\bibinfo {author} {\bibfnamefont {T.}~\bibnamefont
  {Yamaguchi}}, \bibinfo {author} {\bibfnamefont {K.}~\bibnamefont {Sugimoto}},
  \ and\ \bibinfo {author} {\bibfnamefont {Y.}~\bibnamefont {Ohta}},\ }\href
  {\doibase 10.7566/JPSJ.86.043701} {\bibfield  {journal} {\bibinfo  {journal}
  {J. Phys. Soc. Jpn.}\ }\textbf {\bibinfo {volume} {86}},\ \bibinfo {pages}
  {043701} (\bibinfo {year} {2017})}\BibitemShut {NoStop}%
\bibitem [{\citenamefont {Khaliullin}(2013)}]{Khaliullin2013}%
  \BibitemOpen
  \bibfield  {author} {\bibinfo {author} {\bibfnamefont {G.}~\bibnamefont
  {Khaliullin}},\ }\href {\doibase 10.1103/PhysRevLett.111.197201} {\bibfield
  {journal} {\bibinfo  {journal} {Phys. Rev. Lett.}\ }\textbf {\bibinfo
  {volume} {111}},\ \bibinfo {pages} {197201} (\bibinfo {year}
  {2013})}\BibitemShut {NoStop}%
\bibitem [{\citenamefont {Cao}\ \emph {et~al.}(2014)\citenamefont {Cao},
  \citenamefont {Qi}, \citenamefont {Li}, \citenamefont {Terzic}, \citenamefont
  {Yuan}, \citenamefont {DeLong}, \citenamefont {Murthy},\ and\ \citenamefont
  {Kaul}}]{Cao2014}%
  \BibitemOpen
  \bibfield  {author} {\bibinfo {author} {\bibfnamefont {G.}~\bibnamefont
  {Cao}}, \bibinfo {author} {\bibfnamefont {T.~F.}\ \bibnamefont {Qi}},
  \bibinfo {author} {\bibfnamefont {L.}~\bibnamefont {Li}}, \bibinfo {author}
  {\bibfnamefont {J.}~\bibnamefont {Terzic}}, \bibinfo {author} {\bibfnamefont
  {S.~J.}\ \bibnamefont {Yuan}}, \bibinfo {author} {\bibfnamefont {L.~E.}\
  \bibnamefont {DeLong}}, \bibinfo {author} {\bibfnamefont {G.}~\bibnamefont
  {Murthy}}, \ and\ \bibinfo {author} {\bibfnamefont {R.~K.}\ \bibnamefont
  {Kaul}},\ }\href {\doibase 10.1103/PhysRevLett.112.056402} {\bibfield
  {journal} {\bibinfo  {journal} {Phys. Rev. Lett.}\ }\textbf {\bibinfo
  {volume} {112}},\ \bibinfo {pages} {056402} (\bibinfo {year}
  {2014})}\BibitemShut {NoStop}%
\bibitem [{\citenamefont {Jain}\ \emph {et~al.}(2017)\citenamefont {Jain},
  \citenamefont {Krautloher}, \citenamefont {Porras}, \citenamefont {Ryu},
  \citenamefont {Chen}, \citenamefont {Abernathy}, \citenamefont {Park},
  \citenamefont {Ivanov}, \citenamefont {Chaloupka}, \citenamefont
  {Khaliullin}, \citenamefont {Keimer},\ and\ \citenamefont {Kim}}]{Jain2017}%
  \BibitemOpen
  \bibfield  {author} {\bibinfo {author} {\bibfnamefont {A.}~\bibnamefont
  {Jain}}, \bibinfo {author} {\bibfnamefont {M.}~\bibnamefont {Krautloher}},
  \bibinfo {author} {\bibfnamefont {J.}~\bibnamefont {Porras}}, \bibinfo
  {author} {\bibfnamefont {G.~H.}\ \bibnamefont {Ryu}}, \bibinfo {author}
  {\bibfnamefont {D.~P.}\ \bibnamefont {Chen}}, \bibinfo {author}
  {\bibfnamefont {D.~L.}\ \bibnamefont {Abernathy}}, \bibinfo {author}
  {\bibfnamefont {J.~T.}\ \bibnamefont {Park}}, \bibinfo {author}
  {\bibfnamefont {A.}~\bibnamefont {Ivanov}}, \bibinfo {author} {\bibfnamefont
  {J.}~\bibnamefont {Chaloupka}}, \bibinfo {author} {\bibfnamefont
  {G.}~\bibnamefont {Khaliullin}}, \bibinfo {author} {\bibfnamefont
  {B.}~\bibnamefont {Keimer}}, \ and\ \bibinfo {author} {\bibfnamefont {B.~J.}\
  \bibnamefont {Kim}},\ }\href {\doibase 10.1038/nphys4077} {\bibfield
  {journal} {\bibinfo  {journal} {Nat. Phys.}\ }\textbf {\bibinfo {volume}
  {13}},\ \bibinfo {pages} {633} (\bibinfo {year} {2017})}\BibitemShut
  {NoStop}%
\bibitem [{\citenamefont {Tsujimoto}\ \emph {et~al.}(2011)\citenamefont
  {Tsujimoto}, \citenamefont {Li}, \citenamefont {Yamaura}, \citenamefont
  {Matsushita}, \citenamefont {Katsuya}, \citenamefont {Tanaka}, \citenamefont
  {Shirako}, \citenamefont {Akaogi},\ and\ \citenamefont
  {Takayama-Muromachi}}]{Tsujimoto11}%
  \BibitemOpen
  \bibfield  {author} {\bibinfo {author} {\bibfnamefont {Y.}~\bibnamefont
  {Tsujimoto}}, \bibinfo {author} {\bibfnamefont {J.~J.}\ \bibnamefont {Li}},
  \bibinfo {author} {\bibfnamefont {K.}~\bibnamefont {Yamaura}}, \bibinfo
  {author} {\bibfnamefont {Y.}~\bibnamefont {Matsushita}}, \bibinfo {author}
  {\bibfnamefont {Y.}~\bibnamefont {Katsuya}}, \bibinfo {author} {\bibfnamefont
  {M.}~\bibnamefont {Tanaka}}, \bibinfo {author} {\bibfnamefont
  {Y.}~\bibnamefont {Shirako}}, \bibinfo {author} {\bibfnamefont
  {M.}~\bibnamefont {Akaogi}}, \ and\ \bibinfo {author} {\bibfnamefont
  {E.}~\bibnamefont {Takayama-Muromachi}},\ }\href {\doibase
  10.1039/C0CC05482H} {\bibfield  {journal} {\bibinfo  {journal} {Chem.
  Commun.}\ }\textbf {\bibinfo {volume} {47}},\ \bibinfo {pages} {3263}
  (\bibinfo {year} {2011})}\BibitemShut {NoStop}%
\bibitem [{\citenamefont {Tsujimoto}\ \emph {et~al.}(2012)\citenamefont
  {Tsujimoto}, \citenamefont {Sathish}, \citenamefont {Hong}, \citenamefont
  {Oka}, \citenamefont {Azuma}, \citenamefont {Guo}, \citenamefont
  {Matsushita}, \citenamefont {Yamaura},\ and\ \citenamefont
  {Takayama-Muromachi}}]{Tsujimoto12}%
  \BibitemOpen
  \bibfield  {author} {\bibinfo {author} {\bibfnamefont {Y.}~\bibnamefont
  {Tsujimoto}}, \bibinfo {author} {\bibfnamefont {C.~I.}\ \bibnamefont
  {Sathish}}, \bibinfo {author} {\bibfnamefont {K.-P.}\ \bibnamefont {Hong}},
  \bibinfo {author} {\bibfnamefont {K.}~\bibnamefont {Oka}}, \bibinfo {author}
  {\bibfnamefont {M.}~\bibnamefont {Azuma}}, \bibinfo {author} {\bibfnamefont
  {Y.}~\bibnamefont {Guo}}, \bibinfo {author} {\bibfnamefont {Y.}~\bibnamefont
  {Matsushita}}, \bibinfo {author} {\bibfnamefont {K.}~\bibnamefont {Yamaura}},
  \ and\ \bibinfo {author} {\bibfnamefont {E.}~\bibnamefont
  {Takayama-Muromachi}},\ }\href {\doibase 10.1021/ic300116h} {\bibfield
  {journal} {\bibinfo  {journal} {Inorg. Chem.}\ }\textbf {\bibinfo {volume}
  {51}},\ \bibinfo {pages} {4802} (\bibinfo {year} {2012})}\BibitemShut
  {NoStop}%
\bibitem [{\citenamefont {Tsujimoto}\ \emph {et~al.}(2016)\citenamefont
  {Tsujimoto}, \citenamefont {Nakano}, \citenamefont {Ishimatsu}, \citenamefont
  {Mizumaki}, \citenamefont {Kawamura}, \citenamefont {Kawakami}, \citenamefont
  {Matsushita},\ and\ \citenamefont {Yamaura}}]{Tsujimoto16}%
  \BibitemOpen
  \bibfield  {author} {\bibinfo {author} {\bibfnamefont {Y.}~\bibnamefont
  {Tsujimoto}}, \bibinfo {author} {\bibfnamefont {S.}~\bibnamefont {Nakano}},
  \bibinfo {author} {\bibfnamefont {N.}~\bibnamefont {Ishimatsu}}, \bibinfo
  {author} {\bibfnamefont {M.}~\bibnamefont {Mizumaki}}, \bibinfo {author}
  {\bibfnamefont {N.}~\bibnamefont {Kawamura}}, \bibinfo {author}
  {\bibfnamefont {T.}~\bibnamefont {Kawakami}}, \bibinfo {author}
  {\bibfnamefont {Y.}~\bibnamefont {Matsushita}}, \ and\ \bibinfo {author}
  {\bibfnamefont {K.}~\bibnamefont {Yamaura}},\ }\href
  {http://dx.doi.org/10.1038/srep36253} {\bibfield  {journal} {\bibinfo
  {journal} {Sci. Rep.}\ }\textbf {\bibinfo {volume} {6}},\ \bibinfo {pages}
  {36253} (\bibinfo {year} {2016})}\BibitemShut {NoStop}%
\bibitem [{\citenamefont {Afonso}\ \emph {et~al.}(2018)\citenamefont {Afonso},
  \citenamefont {Sotnikov},\ and\ \citenamefont {Kune\v{s}}}]{afonso18}%
  \BibitemOpen
  \bibfield  {author} {\bibinfo {author} {\bibfnamefont {J.~F.}\ \bibnamefont
  {Afonso}}, \bibinfo {author} {\bibfnamefont {A.}~\bibnamefont {Sotnikov}}, \
  and\ \bibinfo {author} {\bibfnamefont {J.}~\bibnamefont {Kune\v{s}}},\ }\href
  {\doibase 10.1088/1361-648X/aab0bb} {\bibfield  {journal} {\bibinfo
  {journal} {J. Phys. Condens. Matter}\ }\textbf {\bibinfo {volume} {30}},\
  \bibinfo {pages} {135603} (\bibinfo {year} {2018})}\BibitemShut {NoStop}%
\bibitem [{\citenamefont {Kohn}\ and\ \citenamefont {Sham}(1965)}]{ksdft}%
  \BibitemOpen
  \bibfield  {author} {\bibinfo {author} {\bibfnamefont {W.}~\bibnamefont
  {Kohn}}\ and\ \bibinfo {author} {\bibfnamefont {L.~J.}\ \bibnamefont
  {Sham}},\ }\href {\doibase 10.1103/PhysRev.140.A1133} {\bibfield  {journal}
  {\bibinfo  {journal} {Phys. Rev.}\ }\textbf {\bibinfo {volume} {140}},\
  \bibinfo {pages} {A1133} (\bibinfo {year} {1965})}\BibitemShut {NoStop}%
\bibitem [{\citenamefont {Hohenberg}\ and\ \citenamefont {Kohn}(1964)}]{hklda}%
  \BibitemOpen
  \bibfield  {author} {\bibinfo {author} {\bibfnamefont {P.}~\bibnamefont
  {Hohenberg}}\ and\ \bibinfo {author} {\bibfnamefont {W.}~\bibnamefont
  {Kohn}},\ }\href {\doibase 10.1103/PhysRev.136.B864} {\bibfield  {journal}
  {\bibinfo  {journal} {Phys. Rev.}\ }\textbf {\bibinfo {volume} {136}},\
  \bibinfo {pages} {B864} (\bibinfo {year} {1964})}\BibitemShut {NoStop}%
\bibitem [{\citenamefont {Blaha}\ \emph {et~al.}(2001)\citenamefont {Blaha},
  \citenamefont {Schwarz}, \citenamefont {Madsen}, \citenamefont {Kvasnicka},\
  and\ \citenamefont {Luitz}}]{wien2k}%
  \BibitemOpen
  \bibfield  {author} {\bibinfo {author} {\bibfnamefont {P.}~\bibnamefont
  {Blaha}}, \bibinfo {author} {\bibfnamefont {K.}~\bibnamefont {Schwarz}},
  \bibinfo {author} {\bibfnamefont {G.~K.~H.}\ \bibnamefont {Madsen}}, \bibinfo
  {author} {\bibfnamefont {D.}~\bibnamefont {Kvasnicka}}, \ and\ \bibinfo
  {author} {\bibfnamefont {J.}~\bibnamefont {Luitz}},\ }\href@noop {} {\emph
  {\bibinfo {title} {{WIEN2K}, {A}n {A}ugmented {P}lane {W}ave + {L}ocal
  {O}rbitals {P}rogram for {C}alculating {C}rystal {P}roperties}}}\ (\bibinfo
  {publisher} {{K}arlheinz Schwarz, Techn. Universit\"{a}t Wien, Austria},\
  \bibinfo {year} {2001})\BibitemShut {NoStop}%
\bibitem [{\citenamefont {Kune\v{s}}\ \emph {et~al.}(2010)\citenamefont
  {Kune\v{s}}, \citenamefont {Arita}, \citenamefont {Wissgott}, \citenamefont
  {Toschi}, \citenamefont {Ikeda},\ and\ \citenamefont {Held}}]{wien2wannier}%
  \BibitemOpen
  \bibfield  {author} {\bibinfo {author} {\bibfnamefont {J.}~\bibnamefont
  {Kune\v{s}}}, \bibinfo {author} {\bibfnamefont {R.}~\bibnamefont {Arita}},
  \bibinfo {author} {\bibfnamefont {P.}~\bibnamefont {Wissgott}}, \bibinfo
  {author} {\bibfnamefont {A.}~\bibnamefont {Toschi}}, \bibinfo {author}
  {\bibfnamefont {H.}~\bibnamefont {Ikeda}}, \ and\ \bibinfo {author}
  {\bibfnamefont {K.}~\bibnamefont {Held}},\ }\href {\doibase
  https://doi.org/10.1016/j.cpc.2010.08.005} {\bibfield  {journal} {\bibinfo
  {journal} {Comp. Phys. Comm.}\ }\textbf {\bibinfo {volume} {181}},\ \bibinfo
  {pages} {1888 } (\bibinfo {year} {2010})}\BibitemShut {NoStop}%
\bibitem [{\citenamefont {Mostofi}\ \emph {et~al.}(2014)\citenamefont
  {Mostofi}, \citenamefont {Yates}, \citenamefont {Pizzi}, \citenamefont {Lee},
  \citenamefont {Souza}, \citenamefont {Vanderbilt},\ and\ \citenamefont
  {Marzari}}]{wannier90}%
  \BibitemOpen
  \bibfield  {author} {\bibinfo {author} {\bibfnamefont {A.~A.}\ \bibnamefont
  {Mostofi}}, \bibinfo {author} {\bibfnamefont {J.~R.}\ \bibnamefont {Yates}},
  \bibinfo {author} {\bibfnamefont {G.}~\bibnamefont {Pizzi}}, \bibinfo
  {author} {\bibfnamefont {Y.-S.}\ \bibnamefont {Lee}}, \bibinfo {author}
  {\bibfnamefont {I.}~\bibnamefont {Souza}}, \bibinfo {author} {\bibfnamefont
  {D.}~\bibnamefont {Vanderbilt}}, \ and\ \bibinfo {author} {\bibfnamefont
  {N.}~\bibnamefont {Marzari}},\ }\href {\doibase
  https://doi.org/10.1016/j.cpc.2014.05.003} {\bibfield  {journal} {\bibinfo
  {journal} {Comp. Phys. Comm.}\ }\textbf {\bibinfo {volume} {185}},\ \bibinfo
  {pages} {2309 } (\bibinfo {year} {2014})}\BibitemShut {NoStop}%
\bibitem [{Note1()}]{Note1}%
  \BibitemOpen
  \bibinfo {note} {The employed amplitude of SOC $\zeta =56$~meV is based on
  the electronic-structure analysis in LaSrCoO$_4${}~\cite
  {afonso18}.}\BibitemShut {Stop}%
\bibitem [{\citenamefont {Karolak}\ \emph {et~al.}(2015)\citenamefont
  {Karolak}, \citenamefont {Izquierdo}, \citenamefont {Molodtsov},\ and\
  \citenamefont {Lichtenstein}}]{karolak15}%
  \BibitemOpen
  \bibfield  {author} {\bibinfo {author} {\bibfnamefont {M.}~\bibnamefont
  {Karolak}}, \bibinfo {author} {\bibfnamefont {M.}~\bibnamefont {Izquierdo}},
  \bibinfo {author} {\bibfnamefont {S.~L.}\ \bibnamefont {Molodtsov}}, \ and\
  \bibinfo {author} {\bibfnamefont {A.~I.}\ \bibnamefont {Lichtenstein}},\
  }\href {\doibase 10.1103/PhysRevLett.115.046401} {\bibfield  {journal}
  {\bibinfo  {journal} {Phys. Rev. Lett.}\ }\textbf {\bibinfo {volume} {115}},\
  \bibinfo {pages} {046401} (\bibinfo {year} {2015})}\BibitemShut {NoStop}%
\bibitem [{\citenamefont {Pavarini}\ \emph {et~al.}(2011)\citenamefont
  {Pavarini}, \citenamefont {Koch}, \citenamefont {Lichtenstein},\ and\
  \citenamefont {Vollhardt}}]{Pavarini1}%
  \BibitemOpen
  \bibfield  {author} {\bibinfo {author} {\bibfnamefont {E.}~\bibnamefont
  {Pavarini}}, \bibinfo {author} {\bibfnamefont {E.}~\bibnamefont {Koch}},
  \bibinfo {author} {\bibfnamefont {A.}~\bibnamefont {Lichtenstein}}, \ and\
  \bibinfo {author} {\bibfnamefont {D.~E.}\ \bibnamefont {Vollhardt}},\ }\href
  {http://juser.fz-juelich.de/record/17645} {\emph {\bibinfo {title} {{T}he
  {LDA}+{DMFT} approach to strongly correlated materials}}},\ \bibinfo {series}
  {Schriften des Forschungszentrums J\"ulich: Modeling and Simulation},
  Vol.~\bibinfo {volume} {1}\ (\bibinfo  {publisher} {Forschungszentrum
  J\"ulich GmbH},\ \bibinfo {year} {2011})\BibitemShut {NoStop}%
\bibitem [{\citenamefont {{Pavarini}}(2014)}]{Pavarini2}%
  \BibitemOpen
  \bibfield  {author} {\bibinfo {author} {\bibfnamefont {E.}~\bibnamefont
  {{Pavarini}}},\ }\enquote {\bibinfo {title} {Electronic structure
  calculations with {LDA+DMFT}},}\ in\ \href {\doibase
  10.1007/978-3-319-06379-9_18} {\emph {\bibinfo {booktitle} {Many-Electron
  Approaches in Physics, Chemistry and Mathematics, Mathematical Physics
  Studies}}},\ \bibinfo {editor} {edited by\ \bibinfo {editor} {\bibfnamefont
  {V.}~\bibnamefont {{Bach}}}\ and\ \bibinfo {editor} {\bibfnamefont
  {L.}~\bibnamefont {{Delle Site}}}}\ (\bibinfo  {publisher} {Springer},\
  \bibinfo {year} {2014})\ p.\ \bibinfo {pages} {321}\BibitemShut {NoStop}%
\bibitem [{\citenamefont {Kune{\v{s}}}(2015)}]{kunes15}%
  \BibitemOpen
  \bibfield  {author} {\bibinfo {author} {\bibfnamefont {J.}~\bibnamefont
  {Kune{\v{s}}}},\ }\href {\doibase 10.1088/0953-8984/27/33/333201} {\bibfield
  {journal} {\bibinfo  {journal} {J. Phys.: Condens. Matter}\ }\textbf
  {\bibinfo {volume} {27}},\ \bibinfo {pages} {333201} (\bibinfo {year}
  {2015})}\BibitemShut {NoStop}%
\bibitem [{\citenamefont {Sommer}\ \emph {et~al.}(2001)\citenamefont {Sommer},
  \citenamefont {Vojta},\ and\ \citenamefont {Becker}}]{svb01}%
  \BibitemOpen
  \bibfield  {author} {\bibinfo {author} {\bibfnamefont {T.}~\bibnamefont
  {Sommer}}, \bibinfo {author} {\bibfnamefont {M.}~\bibnamefont {Vojta}}, \
  and\ \bibinfo {author} {\bibfnamefont {K.}~\bibnamefont {Becker}},\ }\href
  {\doibase 10.1007/s100510170052} {\bibfield  {journal} {\bibinfo  {journal}
  {Eur. Phys. J. B}\ }\textbf {\bibinfo {volume} {23}},\ \bibinfo {pages} {329}
  (\bibinfo {year} {2001})}\BibitemShut {NoStop}%
\bibitem [{\citenamefont {Wallerberger}\ \emph {et~al.}(2019)\citenamefont
  {Wallerberger}, \citenamefont {Hausoel}, \citenamefont {Gunacker},
  \citenamefont {Kowalski}, \citenamefont {Parragh}, \citenamefont {Goth},
  \citenamefont {Held},\ and\ \citenamefont {Sangiovanni}}]{w2dynamics}%
  \BibitemOpen
  \bibfield  {author} {\bibinfo {author} {\bibfnamefont {M.}~\bibnamefont
  {Wallerberger}}, \bibinfo {author} {\bibfnamefont {A.}~\bibnamefont
  {Hausoel}}, \bibinfo {author} {\bibfnamefont {P.}~\bibnamefont {Gunacker}},
  \bibinfo {author} {\bibfnamefont {A.}~\bibnamefont {Kowalski}}, \bibinfo
  {author} {\bibfnamefont {N.}~\bibnamefont {Parragh}}, \bibinfo {author}
  {\bibfnamefont {F.}~\bibnamefont {Goth}}, \bibinfo {author} {\bibfnamefont
  {K.}~\bibnamefont {Held}}, \ and\ \bibinfo {author} {\bibfnamefont
  {G.}~\bibnamefont {Sangiovanni}},\ }\href {\doibase
  10.1016/j.cpc.2018.09.007} {\bibfield  {journal} {\bibinfo  {journal} {Comp.
  Phys. Comm.}\ }\textbf {\bibinfo {volume} {235}},\ \bibinfo {pages} {388 }
  (\bibinfo {year} {2019})}\BibitemShut {NoStop}%
\bibitem [{\citenamefont {Kowalski}\ \emph {et~al.}(2019)\citenamefont
  {Kowalski}, \citenamefont {Hausoel}, \citenamefont {Wallerberger},
  \citenamefont {Gunacker},\ and\ \citenamefont {Sangiovanni}}]{Kowalski18}%
  \BibitemOpen
  \bibfield  {author} {\bibinfo {author} {\bibfnamefont {A.}~\bibnamefont
  {Kowalski}}, \bibinfo {author} {\bibfnamefont {A.}~\bibnamefont {Hausoel}},
  \bibinfo {author} {\bibfnamefont {M.}~\bibnamefont {Wallerberger}}, \bibinfo
  {author} {\bibfnamefont {P.}~\bibnamefont {Gunacker}}, \ and\ \bibinfo
  {author} {\bibfnamefont {G.}~\bibnamefont {Sangiovanni}},\ }\href {\doibase
  10.1103/PhysRevB.99.155112} {\bibfield  {journal} {\bibinfo  {journal} {Phys.
  Rev. B}\ }\textbf {\bibinfo {volume} {99}},\ \bibinfo {pages} {155112}
  (\bibinfo {year} {2019})}\BibitemShut {NoStop}%
\bibitem [{\citenamefont {Geffroy}\ \emph {et~al.}(2019)\citenamefont
  {Geffroy}, \citenamefont {Kaufmann}, \citenamefont {Hariki}, \citenamefont
  {Gunacker}, \citenamefont {Hausoel},\ and\ \citenamefont
  {Kune\ifmmode~\check{s}\else \v{s}\fi{}}}]{Geffroy18}%
  \BibitemOpen
  \bibfield  {author} {\bibinfo {author} {\bibfnamefont {D.}~\bibnamefont
  {Geffroy}}, \bibinfo {author} {\bibfnamefont {J.}~\bibnamefont {Kaufmann}},
  \bibinfo {author} {\bibfnamefont {A.}~\bibnamefont {Hariki}}, \bibinfo
  {author} {\bibfnamefont {P.}~\bibnamefont {Gunacker}}, \bibinfo {author}
  {\bibfnamefont {A.}~\bibnamefont {Hausoel}}, \ and\ \bibinfo {author}
  {\bibfnamefont {J.}~\bibnamefont {Kune\ifmmode~\check{s}\else \v{s}\fi{}}},\
  }\href {\doibase 10.1103/PhysRevLett.122.127601} {\bibfield  {journal}
  {\bibinfo  {journal} {Phys. Rev. Lett.}\ }\textbf {\bibinfo {volume} {122}},\
  \bibinfo {pages} {127601} (\bibinfo {year} {2019})}\BibitemShut {NoStop}%
\bibitem [{\citenamefont {Kune\ifmmode~\check{s}\else \v{s}\fi{}}\ and\
  \citenamefont {K\ifmmode~\check{r}\else \v{r}\fi{}\'apek}(2011)}]{kunes11}%
  \BibitemOpen
  \bibfield  {author} {\bibinfo {author} {\bibfnamefont {J.}~\bibnamefont
  {Kune\ifmmode~\check{s}\else \v{s}\fi{}}}\ and\ \bibinfo {author}
  {\bibfnamefont {V.}~\bibnamefont {K\ifmmode~\check{r}\else
  \v{r}\fi{}\'apek}},\ }\href {\doibase 10.1103/PhysRevLett.106.256401}
  {\bibfield  {journal} {\bibinfo  {journal} {Phys. Rev. Lett.}\ }\textbf
  {\bibinfo {volume} {106}},\ \bibinfo {pages} {256401} (\bibinfo {year}
  {2011})}\BibitemShut {NoStop}%
\bibitem [{\citenamefont {Sotnikov}\ and\ \citenamefont
  {Kune\ifmmode~\check{s}\else \v{s}\fi{}}(2016)}]{sotnikov16}%
  \BibitemOpen
  \bibfield  {author} {\bibinfo {author} {\bibfnamefont {A.}~\bibnamefont
  {Sotnikov}}\ and\ \bibinfo {author} {\bibfnamefont {J.}~\bibnamefont
  {Kune\ifmmode~\check{s}\else \v{s}\fi{}}},\ }\href
  {http://dx.doi.org/10.1038/srep30510} {\bibfield  {journal} {\bibinfo
  {journal} {Sci. Rep.}\ }\textbf {\bibinfo {volume} {6}},\ \bibinfo {pages}
  {30510} (\bibinfo {year} {2016})}\BibitemShut {NoStop}%
\bibitem [{\citenamefont {Blume}\ \emph {et~al.}(1971)\citenamefont {Blume},
  \citenamefont {Emery},\ and\ \citenamefont {Griffiths}}]{BEG}%
  \BibitemOpen
  \bibfield  {author} {\bibinfo {author} {\bibfnamefont {M.}~\bibnamefont
  {Blume}}, \bibinfo {author} {\bibfnamefont {V.~J.}\ \bibnamefont {Emery}}, \
  and\ \bibinfo {author} {\bibfnamefont {R.~B.}\ \bibnamefont {Griffiths}},\
  }\href {\doibase 10.1103/PhysRevA.4.1071} {\bibfield  {journal} {\bibinfo
  {journal} {Phys. Rev. A}\ }\textbf {\bibinfo {volume} {4}},\ \bibinfo {pages}
  {1071} (\bibinfo {year} {1971})}\BibitemShut {NoStop}%
\bibitem [{\citenamefont {Hoston}\ and\ \citenamefont
  {Berker}(1991)}]{hoston91}%
  \BibitemOpen
  \bibfield  {author} {\bibinfo {author} {\bibfnamefont {W.}~\bibnamefont
  {Hoston}}\ and\ \bibinfo {author} {\bibfnamefont {A.~N.}\ \bibnamefont
  {Berker}},\ }\href {\doibase 10.1103/PhysRevLett.67.1027} {\bibfield
  {journal} {\bibinfo  {journal} {Phys. Rev. Lett.}\ }\textbf {\bibinfo
  {volume} {67}},\ \bibinfo {pages} {1027} (\bibinfo {year}
  {1991})}\BibitemShut {NoStop}%
\bibitem [{Note2()}]{Note2}%
  \BibitemOpen
  \bibinfo {note} {In DMFT, the SSO order parameter~$D$ is estimated
  approximately from the electron densities on different $e_g$ and $t_{2g}$
  orbitals with restricting to the subspace of three lowest multiplets LS,
  IS$_1$, and HS$_1$.}\BibitemShut {Stop}%
\bibitem [{\citenamefont {Kn\'{\i}\ifmmode\check{z}\else\v{z}\fi{}ek}\ \emph
  {et~al.}(2009)\citenamefont {Kn\'{\i}\ifmmode\check{z}\else\v{z}\fi{}ek},
  \citenamefont {Jir\'ak}, \citenamefont {Hejtm\'anek}, \citenamefont
  {Nov\'ak},\ and\ \citenamefont {Ku}}]{knizek09}%
  \BibitemOpen
  \bibfield  {author} {\bibinfo {author} {\bibfnamefont {K.}~\bibnamefont
  {Kn\'{\i}\ifmmode\check{z}\else\v{z}\fi{}ek}}, \bibinfo {author}
  {\bibfnamefont {Z.}~\bibnamefont {Jir\'ak}}, \bibinfo {author} {\bibfnamefont
  {J.}~\bibnamefont {Hejtm\'anek}}, \bibinfo {author} {\bibfnamefont
  {P.}~\bibnamefont {Nov\'ak}}, \ and\ \bibinfo {author} {\bibfnamefont
  {W.}~\bibnamefont {Ku}},\ }\href {\doibase 10.1103/PhysRevB.79.014430}
  {\bibfield  {journal} {\bibinfo  {journal} {Phys. Rev. B}\ }\textbf {\bibinfo
  {volume} {79}},\ \bibinfo {pages} {014430} (\bibinfo {year}
  {2009})}\BibitemShut {NoStop}%
\bibitem [{\citenamefont {Bari}\ and\ \citenamefont
  {Sivardi\`ere}(1972)}]{Bari1972}%
  \BibitemOpen
  \bibfield  {author} {\bibinfo {author} {\bibfnamefont {R.~A.}\ \bibnamefont
  {Bari}}\ and\ \bibinfo {author} {\bibfnamefont {J.}~\bibnamefont
  {Sivardi\`ere}},\ }\href {\doibase 10.1103/PhysRevB.5.4466} {\bibfield
  {journal} {\bibinfo  {journal} {Phys. Rev. B}\ }\textbf {\bibinfo {volume}
  {5}},\ \bibinfo {pages} {4466} (\bibinfo {year} {1972})}\BibitemShut
  {NoStop}%
\bibitem [{\citenamefont {{Tomiyasu}}\ \emph {et~al.}(2018)\citenamefont
  {{Tomiyasu}}, \citenamefont {{Nomura}}, \citenamefont {{Kobayashi}},
  \citenamefont {{Ishihara}}, \citenamefont {{Ohira-Kawamura}},\ and\
  \citenamefont {{Kofu}}}]{tomiyasu18pre}%
  \BibitemOpen
  \bibfield  {author} {\bibinfo {author} {\bibfnamefont {K.}~\bibnamefont
  {{Tomiyasu}}}, \bibinfo {author} {\bibfnamefont {T.}~\bibnamefont
  {{Nomura}}}, \bibinfo {author} {\bibfnamefont {Y.}~\bibnamefont
  {{Kobayashi}}}, \bibinfo {author} {\bibfnamefont {S.}~\bibnamefont
  {{Ishihara}}}, \bibinfo {author} {\bibfnamefont {S.}~\bibnamefont
  {{Ohira-Kawamura}}}, \ and\ \bibinfo {author} {\bibfnamefont
  {M.}~\bibnamefont {{Kofu}}},\ }\href@noop {} {\bibfield  {journal} {\bibinfo
  {journal} {ArXiv e-prints}\ } (\bibinfo {year} {2018})},\ \Eprint
  {http://arxiv.org/abs/1808.05888} {arXiv:1808.05888} \BibitemShut {NoStop}%
\bibitem [{\citenamefont {Freeland}\ \emph {et~al.}(2008)\citenamefont
  {Freeland}, \citenamefont {Ma},\ and\ \citenamefont {Shi}}]{freeland08}%
  \BibitemOpen
  \bibfield  {author} {\bibinfo {author} {\bibfnamefont {J.~W.}\ \bibnamefont
  {Freeland}}, \bibinfo {author} {\bibfnamefont {J.~X.}\ \bibnamefont {Ma}}, \
  and\ \bibinfo {author} {\bibfnamefont {J.}~\bibnamefont {Shi}},\ }\href
  {\doibase 10.1063/1.3027063} {\bibfield  {journal} {\bibinfo  {journal}
  {Appl. Phys. Lett.}\ }\textbf {\bibinfo {volume} {93}},\ \bibinfo {pages}
  {212501} (\bibinfo {year} {2008})}\BibitemShut {NoStop}%
\end{thebibliography}%
\end{document}